\documentclass[12pt]{article}
\usepackage{epsfig,amsmath,amssymb,mathrsfs}
\usepackage[utf8]{inputenc}
\usepackage[colorlinks=true,citecolor=blue,,linktocpage=true,linkcolor=blue,urlcolor=black]{hyperref}
\usepackage{mathtools,slashed}
\usepackage{array}

\usepackage[force]{feynmp-auto}
\usepackage{caption}
\usepackage{subcaption}
\usepackage{cleveref}

  {\end{list}}%

\makeatletter
\renewcommand{\section}{\setcounter{equation}{0}\@startsection
 {section}%
 {1}%
 {0pt}%
 {-1\baselineskip}%
 {0.4\baselineskip}%
 {\bfseries\large}}%
\renewcommand{\subsection}{\@startsection
 {subsection}%
 {2}%
 {0pt}%
 {-0.75\baselineskip}%
 {0.2\baselineskip}%
 {\bfseries}}%
\renewcommand{\subsubsection}{\@startsection
 {subsubsection}%
 {3}%
 {0pt}%
 {-0.5\baselineskip}%
 {0.1\baselineskip}%
 {\sc}}%
\makeatother

\DeclareMathAlphabet{\mathpzc}{OT1}{pzc}{m}{it}

\def\be{\begin{equation}}
\def\ee{\end{equation}}
\def\a{\alpha}

\def\g5{\gamma_{5}}

\def\la{\lambda}

\def\m{\mu}
\def\n{\nu}
\def\r{\rho}
\def\s{\sigma}

\def\prslash{{\partial\mkern-9mu/}}

\def\prslash{{\partial\mkern-9mu/}}    

\def\idxn{\int\!\! d^{n}\!x}

\newcommand{\bea}{\begin{eqnarray}}
\newcommand{\eea}{\end{eqnarray}}
\newcommand{\beann}{\begin{eqnarray*}}
\newcommand{\eeann}{\end{eqnarray*}}
\newcommand{\ba}{\begin{array}}
\newcommand{\ea}{\end{array}}

 \def\g {\gamma}

\newcommand{\email}[1]{\href{mailto:#1}{\tt #1}}

\begin{document}

\rightline{\scriptsize{FTUAM-18-5 \quad  IFT-UAM/CSIC-18-014 \quad  FTI/UCM 23-2018}}
\vglue 50pt

\begin{center}

{\LARGE \bf Scattering of fermions in the Yukawa theory coupled to Unimodular Gravity}\\
\vskip 1.0true cm
{\Large S. Gonzalez-Martin}$^{\dagger}$ {\large and} {\Large C. P. Martin}$^{\ddag}$
\\
\vskip .7cm
{
	$^{\dagger}$Departamento de F\'isica Te\'orica and Instituto de F\'{\i}sica Te\'orica (IFT-UAM/CSIC),\\
	Universidad Aut\'onoma de Madrid, Cantoblanco, 28049, Madrid, Spain\\
	\vskip .1cm
	$^{\ddag}${Universidad Complutense de Madrid (UCM), Departamento de Física Teórica I, Facultad de Ciencias Físicas,  Av. Complutense S/N (Ciudad Univ.), 28040 Madrid, Spain}
	
	\vskip .5cm
	\begin{minipage}[l]{.9\textwidth}
		\begin{center}
			\textit{E-mail:}
			\email{sergio.gonzalezm@uam.es},
			\email{carmelop@fis.ucm.es}
			
		\end{center}
	\end{minipage}
}
\end{center}
\thispagestyle{empty}

\begin{abstract}
We compute the lowest order  gravitational UV divergent radiative corrections to the S matrix element of the  $fermion + fermion\rightarrow fermion + fermion$ scattering process in the massive Yukawa theory, coupled either to Unimodular Gravity or to General Relativity. We show that both Unimodular Gravity and General Relativity give rise to the same UV divergent contribution in Dimensional Regularization. This is a nontrivial result, since in the classical action of Unimodular Gravity coupled to the Yukawa theory,  the graviton field does not couple neither to the mass operator nor to the Yukawa operator. This is unlike the General Relativity case. The agreement found points in the direction that Unimodular Gravity and General Relativity give rise to the same quantum theory when coupled to matter, as long as the Cosmological Constant vanishes. Along the way we have come across another unexpected cancellation of UV divergences for both  Unimodular Gravity and General Relativity, resulting in the UV finiteness of the one-loop and $\kappa g^2$ order  of the vertex involving two fermions and one graviton only.

\end{abstract}

\vfill
\clearpage

\section{Introduction}

When quantum  General Relativity  is formulated as an effective quantum field theory there arises a huge disparity between the actual value of the Cosmological Constant and its theoretically expected value. Unimodular gravity supplies \cite{Weinberg:1988cp, Henneaux:1989zc, Smolin:2009ti} a Wilsonian solution to this problem since the vacuum energy does not gravitate: a breach of the equivalence principle is afoot -- see \cite{Percacci:2017fsy}, for a nice review. In this regard, it is to be stressed that when matter is coupled to Unimodular Gravity, no term occurs in the action which couples the classical potential to the graviton field.

As classical theories,  whatever the value of the Cosmological Constant,  Unimodular Gravity and General Relativity are equivalent  \cite{Ellis:2010uc, Ellis:2013eqs, Gao:2014nia, Cho:2014taa} --at least as far as the equations of motion imply \cite{Oda:2016knt, Oda:2016nvc, Chaturvedi:2016fea}. Of course, General Relativity and Unimodular Gravity  are not equivalent as effective quantum field theories for a nonvanishing
Cosmological Constant. And yet, whether that classical equivalence survives the quantization process  is still an open issue, in the case of vanishing Cosmological Constant or, to be more in harmony with Nature, for physical phenomena where the Cosmological Constant can be effectively set to zero. Several papers have been published where this quantum equivalence has been discussed: see Refs. \cite{Alvarez:2005iy, Kluson:2014esa, Saltas:2014cta,Eichhorn:2015bna, Alvarez:2015sba, Bufalo:2015wda, Alvarez:2016uog, Martin:2017ewb, Gonzalez-Martin:2017bvw, Ardon:2017atk}. However, only in three of them  \cite{Martin:2017ewb, Gonzalez-Martin:2017bvw, Gonzalez-Martin:2017fwz} the coupling of Unimodular Gravity with matter has been studied. The fact that in Unimodular Gravity, unlike for General Relativity, the graviton field does not couple in the classical action  to operators such as such $\lambda \phi^4$, the mass terms, or the Yukawa vertex $g\bar{\psi}\psi\phi$, makes one strongly doubt that the equivalence in question holds at the quantum level. Indeed, the coupling of the graviton field with the terms we have just mentioned gives rise to loop contributions which are UV divergent.  Let us stress that the $\lambda \phi^4$, the mass terms and the Yukawa vertex are key ingredients of the Higgs sector of the Standard Model, so to ascertain their effects when coupled to quantum gravity is not a mere academic issue.

It turns out --see  Ref. \cite{Gonzalez-Martin:2017bvw}-- that in the standard multiplicative MS scheme of dimensional regularization, the  gravitational contributions to the beta functions of the quartic, $\lambda$, and Yukawa, $g$, couplings are not the same for Unimodular Gravity as for General Relativity. As discussed in Ref. \cite{Gonzalez-Martin:2017bvw}, this does not necessarily  imply that the UV divergent behaviour of the S matrix elements differ from one theory to the other, for the aforementioned gravitational corrections have no intrinsic physical meaning. Notice that if the discrepancy  between the gravitational corrections to the beta functions we have just mentioned would necessarily imply unequal UV divergent behaviour of the S matrix elements of Unimodular Gravity and General Relativity coupled to matter, then, one would be entitled to conclude that Unimodular Gravity and General Relativity are not equivalent at the quantum level, at least when interacting with matter. However, what the discrepancy does say is that it is far from clear that Unimodular Gravity and General Relativity agree at
the quantum level as they do at the classical level. It is then a pressing matter to check whether the UV divergent behaviour of the gravitational corrections to the S matrix elements of massive  scalar and fermion particles  for Unimodular Gravity agrees with those coming from General Relativity.

In Ref. \cite{Gonzalez-Martin:2017fwz}, the lowest order UV divergent gravitational radiative corrections to the S matrix  of the scattering of two scalar particles going into two scalar particles was computed, both in  Unimodular Gravity and General Relativity coupled to the massive $\lambda \phi^4$ theory. The outcome was that the total contribution in the Unimodular Gravity case is the same as in the General Relativity case,  although the contribution coming from each individual Feynman diagram is not the same for Unimodular Gravity as for General Relativity.

The purpose of this paper is to compute the one-loop and $g^2\kappa^2 $ order --ie, the lowest order radiative contributions-- UV divergent  contributions to the S matrix element of the scattering process
$fermion + fermion \rightarrow fermion + fermion$ in the massive Yukawa theory coupled either to  General Relativity or to Unimodular Gravity, when the Cosmological Constant is set to zero. We shall show that such  UV divergent behaviour is  the same in Unimodular Gravity case as in the General Relativity instance,  notwithstanding the fact that this equivalence does not hold  Feynman diagram by Feynman diagram. This result is not trivial since  Unimodular Gravity does not couple neither to the mass terms nor the Yukawa vertex $g\bar{\psi}\psi\phi$ in the classical action, besides the fact that they have different gauge symmetries. Our result provides further evidence that Unimodular Gravity and General Relativity are same quantum theory  for zero Cosmological Constant.

The lay out of this paper is as follows. In section 2 we display the relevant formulae that are needed to carry out the computations in Section 3.
Section 3 is devoted to the computation of the one-loop and $g^2\kappa^2$ order,  UV divergent contributions to the S matrix element of the scattering $fermion + fermion \rightarrow fermion + fermion$. Finally, we have a section to discuss the results presented in the paper.

\section{ Yukawa theory coupled to Gravity}

In this section we shall just display the classical actions of the Yukawa theory coupled to a gravitational field as described by General Relativity and Unimodular Gravity. We shall also display the graviton free propagator in each case.

\subsection{Yukawa theory coupled to General Relativity}

Let $e^{\mu}_a$ be the vielbein, $e^{\mu}_a e^{\nu}_b g_{\mu\nu} = \eta_{ab}$, for the Lorentzian metric $g_{\mu\nu}$, $\eta_{ab}=(+,-,-,-)$. Let $\gamma^a$ denote the Dirac matrices:
$[\gamma^a,\gamma^b] =\eta^{a b}$. The torsion-free spin connection $\omega_{\mu}$ is defined,
\begin{equation*}
\omega_{\mu}=\frac{1}{8}[\gamma^b,\gamma^c]e_{\nu b}\bigtriangledown_{\mu}e^{\nu}_c,
\end{equation*}
 where $\bigtriangledown_{\mu}$ is the covariant derivative as given by the Christoffel symbols. Let  $\psi$, denote a spinor field in spacetime, its covariant derivative being given by
\begin{equation*}
D_{\mu}\psi = (\partial_{\mu}+\omega_{\mu})\psi .
\end{equation*}

The classical action of General relativity coupled to the Yukawa theory reads
\begin{equation}
\begin{array}{l}
{S_{\text{GR-Yukawa}}\;=\;S_{\text{EH}}\,+\,S^{\text{(GR)}}_\text{Yukawa}},\\[8pt]
{\displaystyle S_\text{EH}=-\dfrac{2}{\kappa^2}\idxn\;\sqrt{-g} R[g_{\mu\nu}]},\\[8pt]
{\displaystyle S^\text{(GR)}_\text{Yukawa}\,=\,\idxn\,\sqrt{-g}\,\Big[\bar{\psi}\big(i e^{\mu}_a\,\gamma^a D_{\mu}\psi-m\psi\big)+
\dfrac{1}{2}g^{\mu\nu}\partial_\mu\phi\partial_\nu\phi-\dfrac{1}{2}M^2\phi^2- g\,\bar{\psi}\psi\phi}.
\end{array}
\label{classactionGR}
\end{equation}
where $\kappa^2=32\pi G$ and $R[g_{\mu\nu}]$ is the scalar curvature for the metric $g_{\mu\nu}$.

Using the standard splitting
\begin{equation}
g_{\mu\nu}=\eta_{\mu\nu}\,+\,\kappa h_{\mu\nu},
\label{gsplittingGR}
\end{equation}
and the generalized de Donder gauge-fixing term
\begin{equation*}
\idxn\,\alpha (\partial^\mu h_{\mu\nu}-\partial_\nu h)^2,\quad h=h_{\mu\nu}\eta^{\mu\nu},
 \end{equation*}
which depends on the gauge parameter $\alpha$, one obtains the following free propagator of the graviton field $h_{\mu\nu}$,
\begin{equation}
\begin{array}{l}
{\langle h_{\mu\nu}(k)h_{\rho\sigma}(-k)\rangle=
\dfrac{i}{2k^2}\left(\eta_{\m\s}\eta_{\n\r}+\eta_{\m\r}\eta_{\n\s}-\eta_{\m\n}\eta_{\r\s}\right)}-\\[8pt]
{\phantom{\langle h_{\mu\nu}(k)h_{\rho\sigma}(-k)\rangle=}
-\dfrac{i}{(k^2)^2}\,(\dfrac{1}{2}+\alpha)\left(\eta_{\m\r}k_\n k_\s+\eta_{\m\s}k_\n k_\r+\eta_{\n\r}k_\m k_\s+\eta_{\n\s}k_\m k_\r\right)}.
\end{array}
\label{GRpropagator}
\end{equation}	
where $\eta^{\mu\nu}$ denotes the Minkowski, $(+,-,-,-)$, metric.

Up to first order in $\kappa$, $S^{(GR)}_{Yukawa}$ in (\ref{classactionGR}) is given by
\begin{equation}
S^\text{(GR)}_{\text{Yukawa}}=\,\idxn\,\Big[\bar{\psi}\big(i\prslash-m\big)\psi+\dfrac{1}{2}\partial_\mu\phi\partial^\mu\phi-\dfrac{1}{2}M^2\phi^2-g\bar{\psi}\psi\phi-\dfrac{\kappa}{2}\,T^{\mu\nu}h_{\mu\nu}\Big]+O(\kappa^2),
\label{Sphi4GR}
\end{equation}
where
 \begin{equation}
 \begin{array}{l}
 {T^{\mu\nu}=\dfrac{i}{4}\bar{\psi}\big(\gamma^\mu\overrightarrow{\partial}^\nu+ \gamma^\nu\overrightarrow{\partial}^\mu\big)\psi
- \dfrac{i}{4}\bar{\psi}\big(\gamma^\mu\overleftarrow{\partial}^\nu+ \gamma^\nu\overleftarrow{\partial}^\mu\big)\psi+
 \partial^{\mu}\phi\partial^{\nu}\phi}\\[4pt]
 {\phantom{T^{\mu\nu}=}-\eta^{\mu\nu}\big(\dfrac{1}{2}\bar{\psi}(i\overrightarrow{\prslash} -m)\psi- \dfrac{1}{2}\bar{\psi}(i\overleftarrow{\prslash} +m)\psi+\dfrac{1}{2}\partial_\la\phi\partial^\la\phi-\dfrac{1}{2}M^2\phi^2-g\bar{\psi}\psi\phi\big).}
 \end{array}
 \label{EMtensor}
 \end{equation}
In \eqref{Sphi4GR}, contractions are carried out with $\eta_{\mu\nu}$.

\subsection{Yukawa theory coupled to Unimodular Gravity}

Let $\hat{g}_{\mu\nu}$ denote the Unimodular --ie, with determinant equal to -1-- metric of the $n$ dimensional spacetime manifold. We shall assume the mostly minus signature for the metric. Then,
the classical action of the Yukawa theory coupled to Unimodular Gravity reads
\begin{equation}
\begin{array}{l}
{S_\text{UG-Yukawa}\;=\;S_{UG}\,+\,S^{(\text{UG})}_\text{Yukawa}},\\[8pt]
{\displaystyle S_{UG}=-\dfrac{2}{\kappa^2}\idxn\; R[\hat{g}_{\mu\nu}]},\\[8pt]
{\displaystyle S^{(\text{UG})}_\text{Yukawa}\,=\,\idxn\,\Big[\bar{\psi}\big(i \hat{e}^{\mu}_a\,\gamma^a \hat{D}_{\mu}\psi-m\psi\big)+
\dfrac{1}{2}\hat{g}^{\mu\nu}\partial_\mu\phi\partial_\nu\phi-\dfrac{1}{2}M^2\phi^2- g\,\bar{\psi}\psi\phi},
\end{array}
\label{classactionUG}
\end{equation}
where $\kappa^2=32\pi G$, $R[\hat{g}_{\mu\nu}]$ is the scalar curvature for the unimodular metric, $\hat{e}^{\mu}_a$ is the vielbein, $\hat{e}^{\mu}_a \hat{e}^{\nu}_b \hat{g}_{\mu\nu} = \eta_{ab}$ for the metric $\hat{g}_{\mu\nu}$, $\eta_{ab}=(+,-,-,-)$, $\gamma^a$ denote the Dirac matrices: $[\gamma^a,\gamma^b] =\eta^{ab}$ and $\hat{D}_{\mu}=\partial_\mu+\hat{\omega}_\mu$ is the Dirac operator  for the torsion-free spin connection
\begin{equation*}
\hat{\omega}_{\mu}=\frac{1}{8}[\gamma^b,\gamma^c]\hat{e}_{\nu b}\hat{\bigtriangledown}_{\mu}\hat{e}^{\nu}_c.
\end{equation*}
$\hat{\bigtriangledown}_{\mu}$ is the covariant derivative as given by the Christoffel symbols of $\hat{g}_{\mu\nu}$

To quantize the theory we shall follow Refs.~\cite{Alvarez:2005iy, Alvarez:2015sba, Alvarez:2006uu} and  introduce an unconstrained fictitious metric, $\hat{g}_{\mu\n}$, thus
\begin{equation}
\hat{g}_{\mu\nu}=(-g)^{-1/n}\,g_{\mu\nu},
\label{fictitious}
\end{equation}
where $g$ is the determinant of $g_{\mu\nu}$. Next, we shall express the action in (\ref{classactionUG}) in terms of the fictitious metric $\hat{g}_{\mu\nu}$ by using  (\ref{fictitious}), then,  we shall
split  $g_{\mu\nu}$ as in \eqref{gsplittingGR}
\begin{equation*}
g_{\mu\nu}=\eta_{\mu\nu}\,+\,\kappa h_{\mu\nu};
\end{equation*}
and, finally, we shall define the path integral  by integration over $h_{\mu\nu}$ and the matter fields, once an appropriate BRS invariant action has been constructed.

Since our computations will always involve the matter fields $\bar{\psi}$, $\psi$ and $\phi$, and will be of order $\kappa^2$, we shall only need --as will become clear in the sequel-- the free propagator of $h_{\mu\nu}$,  $\langle h_{\mu\nu}(k)h_{\rho\sigma}(-k)\rangle$,  and the expansion of $S^{(\text{UG})}_\text{Yukawa}$ up to first order in $\kappa$. Using the gauge-fixing procedure discussed in Ref.~\cite{Alvarez:2015sba}, one obtains
\begin{equation}
\begin{array}{l}
{\langle h_{\mu\nu}(k)h_{\rho\sigma}(-k)\rangle=
\dfrac{i}{2k^2}\left(\eta_{\m\s}\eta_{\n\r}+\eta_{\m\r}\eta_{\n\s}\right)-\dfrac{i}{k^2}\dfrac{\a^2n^2-n+2}{\a^2 n^2(n-2)}\eta_{\m\n}\eta_{\r\s}}+\\[8pt]
{\phantom{\langle h_{\mu\nu}(k)h_{\rho\sigma}(-k)\rangle=}
+\dfrac{2i}{n-2}\left(\dfrac{k_\r k_\s \eta_{\m\n}}{(k^2)^2}+\dfrac{k_\m k_\n \eta_{\r\s}}{(k^2)^2}\right)-\dfrac{2in}{n-2}\dfrac{k_{\m}k_{\n}k_{\r}k_{\s}}{(k^2)^3}.
}
\end{array}
\label{propagatorug}
\end{equation}	
The expansion of $S^{(\text{UG})}_\text{Yukawa}$ in powers of $\kappa$ reads
\begin{equation}
S^{(\text{UG})}_\text{Yukawa}=\,\idxn\,\Big[\bar{\psi}\big(i\prslash-m\big)\psi+\dfrac{1}{2}\partial_\mu\phi\partial^\mu\phi-\dfrac{1}{2}M^2\phi^2-g\bar{\psi}\psi\phi-\dfrac{\kappa}{2}\,T^{\mu\nu}\hat{h}_{\mu\nu}\Big]+O(\kappa^2),
\label{Sphi4expan}
\end{equation}
where $\hat{h}_{\mu\nu}=h_{\mu\nu}-\dfrac{1}{n}h$ --with $h=\eta_{\mu\nu}h^{\mu\nu}$-- is the traceless part of $h_{\mu\nu}$, and $T^{\mu\nu}$ is given in (\ref{EMtensor}). Again, the contractions in \eqref{Sphi4expan} are carried out with the help of $\eta_{\mu\nu}$.

Let us point out that the term in $T^{\mu\nu}$ which is proportional to $\eta^{\mu\nu}$ does not actually contribute to  $T^{\mu\nu}\hat{h}_{\mu\nu}$, since $\hat{h}_{\mu\nu}$ is traceless. In terms of Feynman diagrams, this can be stated by saying that the $\eta^{\mu\nu}$ part of $T^{\mu\nu}$ will never
contribute to a given diagram since it will always be contracted with a free propagator involving $\hat{h}_{\mu\nu}$. This is not what happens in the General Relativity case and makes the agreement between  General Relativity coupled to the Yukawa theory and Unimodular Gravity coupled to the latter a non-trivial issue already at one-loop.

We shall use the correlation function $\langle\hat{h}_{\mu\nu}(k)\hat{h}_{\rho\sigma}(-k)\rangle$, which can be easily obtained from (\ref{propagatorug}). It reads
\begin{equation}
\begin{array}{ll}
{\langle\hat{h}_{\mu\nu}(k)\hat{h}_{\rho\sigma}(-k)\rangle=\dfrac{i}{2k^2}\left(\eta_{\m\s}\eta_{\n\r}+\eta_{\m\r}\eta_{\n\s}-\dfrac{2}{n-2}\eta_{\m\n}\eta_{\r\s}\right)+}\\[8pt]
{\phantom{\langle\hat{h}_{\mu\nu}(k)\hat{h}_{\rho\sigma}(-k)\rangle=}+
\dfrac{2i}{n-2}\dfrac{k_{\mu}k_{\nu}\eta_{\rho\sigma}+k_{\rho}k_{\sigma}\eta_{\mu\nu}}{(k^2)^2}-
\dfrac{2in}{n-2}\dfrac{k_{\mu}k_{\nu}k_{\rho}k_{\sigma}}{(k^2)^3}.}
\label{tracelessprop}
\end{array}
\end{equation}

\section{The $fermion + fermion \rightarrow fermion + fermion$ scattering at one-loop and $g^2\kappa^2$ order}

The purpose of this section is to work out the one-loop, and $g^2\kappa^2$ order, UV divergent contribution, coming from General Relativity and Unimodular Gravity, to the dimensionally regularized S matrix element of the  $fermion + fermion \rightarrow fermion + fermion$ scattering process.

\subsection{The General Relativity case}

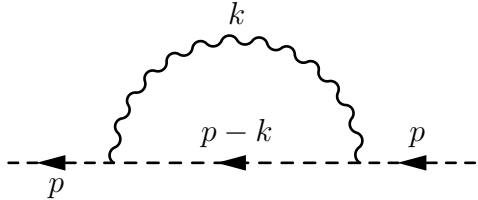
\begin{figure}[h]
	\centering
	\begin{fmffile}{scalarprop2}
		\begin{fmfgraph*}(180,90)
			\fmfleft{i}
			\fmfright{o}
			\fmf{scalar,tension=3,label=$p$}{v1,i}
			\fmf{scalar,tension=0.7,label=$p-k$}{v2,v1}
			\fmf{scalar,tension=3,label=$p$}{o,v2}
			\fmf{wiggly,left,tension=0.7,label=$k$}{v1,v2}
				\end{fmfgraph*}
	\end{fmffile}\vspace{-1cm}
	\caption{Scalar propagator.}
	\label{figure1}
\end{figure}

\begin{figure}[h]
	\centering
	\begin{fmffile}{fermionprop2}
		\begin{fmfgraph*}(180,90)
			\fmfleft{i}
			\fmfright{o}
			\fmf{fermion,tension=3,label=$p$}{v1,i}
			\fmf{fermion,tension=0.7,label=$p-k$}{v2,v1}
			\fmf{fermion,tension=3,label=$p$}{o,v2}
			\fmf{wiggly,right,tension=0.7,label=$k$}{v2,v1}
		\end{fmfgraph*}\vspace{-1cm}
	\end{fmffile}
	\caption{Fermion propagator.}
	\label{figure2}
\end{figure}
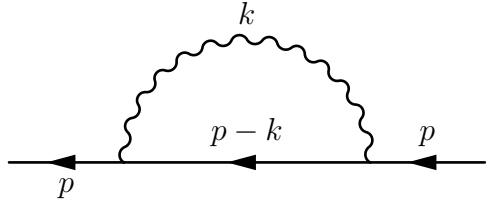

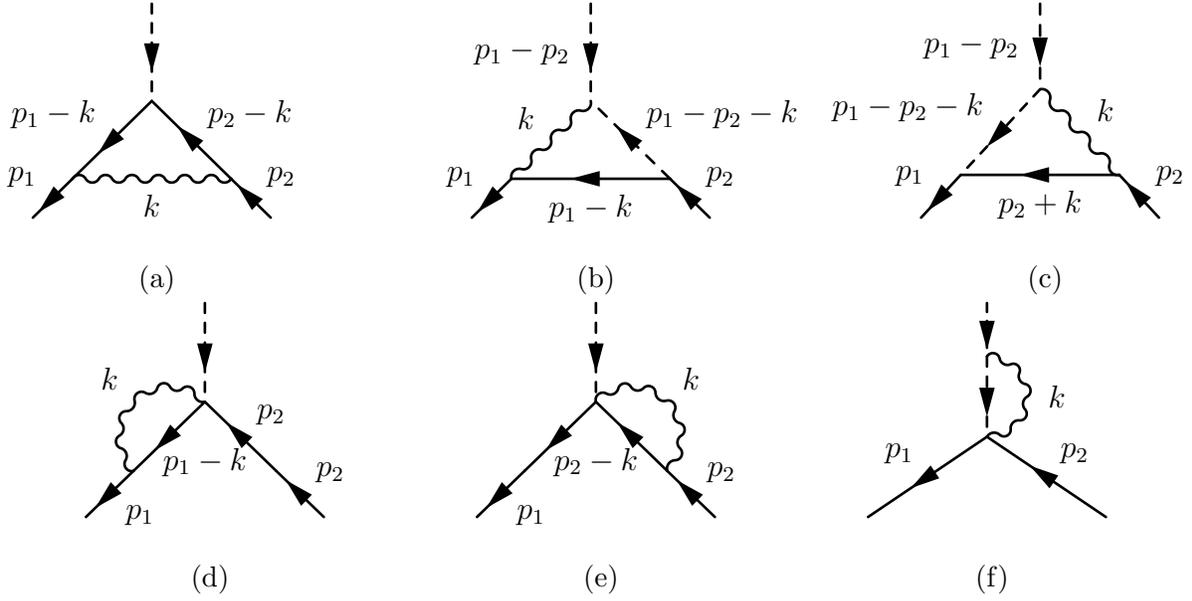
\begin{figure}[h!]
	\centering
	\vspace{1cm}
	\begin{subfigure}{0.3\linewidth}
		\centering
		\begin{fmffile}{yukawa1}
			\begin{fmfgraph*}(90,90)
				\fmfbottom{i,o}
				\fmftop{t}
				\fmf{scalar,tension=0.8}{t,v1}
				\fmf{fermion,tension=0.5,label=$p_1-k$,l.side=right}{v1,v3}
				\fmf{fermion,tension=1,label=$p_1$}{v3,i}
				\fmf{fermion,tension=1,label=$p_2$,l.side=right}{o,v2}
				\fmf{fermion,tension=0.5,label=$p_2-k$}{v2,v1}
				\fmf{wiggly,tension=0,label=$k$,l.side=left}{v2,v3}
			\end{fmfgraph*}
		\end{fmffile}
		\caption{}
	\end{subfigure}\hfill
	\begin{subfigure}{0.3\linewidth}
		\centering
		\begin{fmffile}{yukawa2}
			\begin{fmfgraph*}(90,90)
				\fmfbottom{i,o}
				\fmftop{t}
				\fmf{scalar,tension=0.8,label=$p_1-p_2$}{t,v1}
				\fmf{wiggly,tension=0.5,label=$k$,l.side=right}{v1,v3}
				\fmf{fermion,tension=1,label=$p_1$}{v3,i}
				\fmf{fermion,tension=1,label=$p_2$,l.side=right}{o,v2}
				\fmf{scalar,tension=0.5,label=$p_1-p_2-k$}{v2,v1}
				\fmf{fermion,tension=0,label=$p_1-k$,l.side=left}{v2,v3}
			\end{fmfgraph*}
		\end{fmffile}
		\caption{}
	\end{subfigure}
	\hfill
	\begin{subfigure}{0.3\linewidth}
		\centering
		\begin{fmffile}{yukawa3}
			\begin{fmfgraph*}(90,90)
				\fmfbottom{i,o}
				\fmftop{t}
				\fmf{scalar,tension=1,label=$p_1-p_2$}{t,v1}
				\fmf{scalar,tension=0.5,label=$~~p_1-p_2-k$,l.side=right}{v1,v3}
				\fmf{fermion,tension=1,label=$p_1$}{v3,i}
				\fmf{fermion,tension=1,label=$p_2$,l.side=right}{o,v2}
				\fmf{wiggly,tension=0.5,label=$k$,l.side=right}{v2,v1}
				\fmf{fermion,plain,tension=0,label=$p_2+k$,l.side=left}{v2,v3}
			\end{fmfgraph*}
		\end{fmffile}
		\caption{}
	\end{subfigure}
	
	\begin{subfigure}{0.3\linewidth}
		\centering
		\begin{fmffile}{yukawa4}
			\begin{fmfgraph*}(90,90)
				\fmfbottom{i,o}
				\fmftop{t}
				\fmf{scalar,tension=1.4}{t,v1}
				\fmf{fermion,tension=1,label=\hspace{-0.3cm}$p_1-k$,l.side=left}{v1,v3}
				\fmf{fermion,tension=1.5,label=$p_1$}{v3,i}
				\fmf{fermion,tension=1.5,label=$p_2$}{o,v2}
				\fmf{fermion,tension=1,label=$p_2$}{v2,v1}
				\fmf{wiggly,right,tension=0.0,label=$k$,l.side=right}{v1,v3}
			\end{fmfgraph*}
		\end{fmffile}
		\caption{}
	\end{subfigure}
	\begin{subfigure}{0.3\linewidth}
		\centering
		\begin{fmffile}{yukawa5}
			\begin{fmfgraph*}(90,90)
				\fmfbottom{i,o}
				\fmftop{t}
				\fmf{scalar,tension=1.4}{t,v1}
				\fmf{fermion,tension=1,label=\hspace{-0.3cm}$p_2-k$,l.side=left}{v1,v3}
				\fmf{fermion,tension=1.5,label=$p_1$}{v3,i}
				\fmf{fermion,tension=1.5,label=$p_2$}{o,v2}
				\fmf{fermion,tension=1}{v2,v1}
				\fmf{wiggly,right,tension=0.0,label=$k$}{v2,v1}
			\end{fmfgraph*}
		\end{fmffile}
		\caption{}
	\end{subfigure}
	\begin{subfigure}{0.3\linewidth}
		\centering
		\begin{fmffile}{yukawa6}
			\begin{fmfgraph*}(90,90)
				\fmfbottom{i,o}
				\fmftop{t}
				\fmf{scalar,tension=3}{t,v2,v1}
				\fmf{fermion,tension=1,label=$p_1$}{v1,i}
				\fmf{fermion,tension=1,label=$p_2$,l.side=right}{o,v1}
				\fmf{wiggly,right,tension=-1,label=$k$}{v1,v2}
			\end{fmfgraph*}
		\end{fmffile}
		\caption{}
	\end{subfigure}
	\caption{Vertices of order $g\kappa^2$.}
	\label{figure3}
\end{figure}

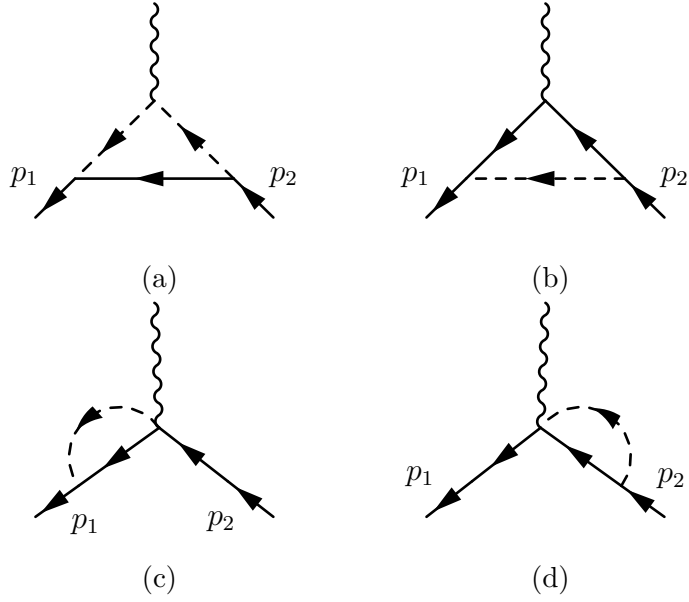
\begin{figure}[h!]
	\centering
	
	\begin{subfigure}{0.3\linewidth}
		\centering
		\begin{fmffile}{yukawa11}
			\begin{fmfgraph*}(90,90)
				\fmfbottom{i,o}
				\fmftop{t}
				\fmf{wiggly,tension=0.8}{t,v1}
				\fmf{scalar,tension=0.5}{v1,v3}
				\fmf{fermion,tension=1,label=$p_1$}{v3,i}
				\fmf{fermion,tension=1,label=$p_2$,l.side=right}{o,v2}
				\fmf{scalar,tension=0.5}{v2,v1}
				\fmf{fermion,tension=0,l.side=left}{v2,v3}
			\end{fmfgraph*}
		\end{fmffile}
		\caption{}
	\end{subfigure}
	\begin{subfigure}{0.3\linewidth}
		\centering
		\begin{fmffile}{yukawa12}
			\begin{fmfgraph*}(90,90)
				\fmfbottom{i,o}
				\fmftop{t}
				\fmf{wiggly,tension=0.8}{t,v1}
				\fmf{fermion,tension=0.5}{v1,v3}
				\fmf{fermion,tension=1,label=$p_1$}{v3,i}
				\fmf{fermion,tension=1,label=$p_2$,l.side=right}{o,v2}
				\fmf{fermion,tension=0.5}{v2,v1}
				\fmf{scalar,tension=0,l.side=left}{v2,v3}
			\end{fmfgraph*}
		\end{fmffile}
		\caption{}
	\end{subfigure}
	
	\begin{subfigure}{0.3\linewidth}
		\centering
		\begin{fmffile}{yukawa13}
			\begin{fmfgraph*}(90,90)
				\fmfbottom{i,o}
				\fmftop{t}
				\fmf{wiggly,tension=0.8}{t,v1}
				\fmf{fermion,tension=1}{v1,v3}
				\fmf{fermion,tension=1.5,label=$p_1$}{v3,i}
				\fmf{fermion,tension=1.5,label=$p_2$}{o,v2}
				\fmf{fermion,tension=1}{v2,v1}
				\fmf{scalar,right,tension=-0.2}{v1,v3}
			\end{fmfgraph*}
		\end{fmffile}
		\caption{}
	\end{subfigure}
	\begin{subfigure}{0.3\linewidth}
		\centering
		\begin{fmffile}{yukawa14}
			\begin{fmfgraph*}(90,90)
				\fmfbottom{i,o}
				\fmftop{t}
				\fmf{wiggly,tension=0.8}{t,v1}
				\fmf{fermion,tension=1}{v1,v3}
				\fmf{fermion,tension=1.5,label=$p_1$}{v3,i}
				\fmf{fermion,tension=1.5,label=$p_2$}{o,v2}
				\fmf{fermion,tension=1}{v2,v1}
				\fmf{scalar,right,tension=-0.2,l.side=right}{v2,v1}
			\end{fmfgraph*}
		\end{fmffile}
		\caption{}
	\end{subfigure}
	\caption{Vertices of order $g^2\kappa$.}
	\label{figure4}
\end{figure}

To work out the UV divergent contribution in question to $fermion + fermion \rightarrow fermion + fermion$, we shall need the UV divergent contributions coming from the 1PI diagramas in Figures \ref{figure1} to \ref{figure4}. These
contributions read in the General Relativity case
\begin{equation}
\begin{array}{l}
{i\Gamma^{(GR)}_{\phi\phi}(p^2,\kappa)= \Big(\dfrac{1}{16\pi^2\epsilon}\Big)\left[1+ \left(\dfrac{1}{2}+\alpha\right)\right]\kappa^2 M^2 (p^2-M^2),}\\[4pt]
{i\Gamma^{(GR)}_{\psi\bar{\psi}}(p;\kappa)=\Big(\dfrac{-i}{16\pi^2\epsilon}\Big)\kappa^2 \Big[\left(\dfrac{3}{8}p^2 m-\dfrac{1}{8}p^2\slashed{p}+\dfrac{1}{4}m^2(\slashed{p}-m)\right)}+\\[4pt]
{\phantom{i\Gamma^{(GR)}_{\psi\bar{\psi}}(p;\kappa)=\quad}
+\left(\dfrac{1}{2}+\alpha\right)\left(\dfrac{3}{4}p^2 m-\dfrac{19}{16}m^3+\left(\dfrac{29}{32}m^2-\dfrac{15}{32}p^2\right )\slashed{p}\right)\Big],}\\[4pt]
{i\Gamma^{(GR)}_{\psi\bar{\psi}\phi}(p_1,p_2;\kappa)=g\kappa^2 \left(-\dfrac{i}{16\pi^2\epsilon}\right)\Big\{\Big( -\dfrac{1}{4}M^2-\dfrac{3}{4}m^2+\dfrac{1}{16}(p_1+p_2)^2+\dfrac{1}{4}m(\slashed{p}_1+\slashed{p}_2)+\dfrac{1}{8}\slashed{p}_1\slashed{p}_2\Big)}+\\[4pt]
{\quad\quad\quad\quad\quad\quad\quad	+ \Big(\dfrac{1}{2}+\a\Big)\left(-m_\phi^2-\dfrac{57}{16}m^2+\dfrac{47}{32}(p_1^2+p_2^2)-\dfrac{13}{8}p_1\cdot p_2 +m (\slashed{p}_1+\slashed{p}_2)-\dfrac{9}{16}\slashed{p}_1\slashed{p}_2\right)\Bigg\},}\\[4pt]
{i\Gamma^{(GR)}_{\psi\bar{\psi} h_{\mu\nu}}(p_1,p_2;\kappa)=\kappa g^2\Big(\dfrac{-i}{16\pi^2\epsilon}\Big)\left\{\left[\dfrac{1}{8}(\slashed{p}_1+\slashed{p}_2)+\dfrac{1}{2}m\right]\eta^{\mu\nu}-\dfrac{1}{16}(p_1+p_2)^{\mu}\gamma^\nu-\dfrac{1}{16}(p_1+p_2)^{\nu}\gamma^\mu\right\}.
}
\end{array}
\label{contributionsone}
 \end{equation}
where $i\Gamma^{(GR)}_{\phi\phi}(p^2,\kappa)$ and  $i\Gamma^{(GR)}_{\psi\bar{\psi}}(p;\kappa)$ are  given by the diagrams in Figure \ref{figure1} and Figure \ref{figure2} , respectively. The wavy line stands for the graviton propagator in
(\ref {GRpropagator}). $i\Gamma^{(GR)}_{\psi\bar{\psi}\phi}(p_1,p_2;\kappa)$ is obtained by adding up all the contributions  the diagrams in Figure \ref{figure3} give rise to. The sum of the diagrams in Figure \ref{figure4} yields $i\Gamma^{(GR)}_{\psi\bar{\psi} h_{\mu\nu}}(p_1,p_2;\kappa)$.\vspace{1em}

To define the S matrix elements it is necessary to express the bare masses $M$ and $m$ in terms of the corresponding physical masses --ie, the poles of the propagators-- $M_{\phi}$ and $m_{\psi}$. This is
accomplished by using the following formulae
\begin{equation}
\begin{array}{l}
{m=Z_m\,m_{\psi},\quad M^2=Z_M\,M^2_{\phi},}\\[4pt]
{Z_m = 1-\Sigma_1(p^2=m_{\psi}^2)-\Sigma_2(p^2=m_{\psi}^2),\quad Z_M=1-\dfrac{i}{M_\phi^2}\Gamma_{\phi\phi}(p^2=M_\phi^2),}\\[4pt]
{i\Gamma_{\psi\bar{\psi}}(p,\kappa)=i\Gamma^{(NG)}_{\psi\bar{\psi}}(p)+ i\Gamma^{(G)}_{\psi\bar{\psi},}(p;\kappa)},\\[4pt]
{i\Gamma_{\phi \phi}(p;\kappa)=i\Gamma^{(NG)}_{\phi\phi}(p)+ i\Gamma^{(G)}_{\phi\phi}(p;\kappa),}\\[4pt]
{i\Gamma^{(NG)}_{\psi\bar{\psi}}(p,\kappa)=(\slashed{p}-m)\Sigma^{(NG)}_2(p^2)+m\Sigma^{(NG)}_1 (p^2),}\\[4pt]
{i\Gamma^{(G)}_{\psi\bar{\psi}}(p,\kappa)=(\slashed{p}-m)\Sigma^{(G)}_2(p^2)+m\Sigma^{(G)}_1 (p^2),}
\end{array}
\label{zetamass}
\end{equation}
where the superscript $G$ stand for gravitational --those from General Relativity  in the current subsection-- contributions, given in (\ref{contributionsone}), and the superscript $NG$ denote the corresponding contributions in absence of gravity, whose actual vaalues are not needed in this paper.

To obtain the S matrix elements from the Green functions of the fields, it is also convenient to introduce  renormalized fields $\phi_R$ and $\psi_R$, so that the Laurent expansion of their propagators,
$\langle\phi_R (p)\phi_R (-p)\rangle$ and $\langle\psi_R (p)\bar{\psi}_R (p)\rangle$ , around the mass shell read

\begin{equation*}
\begin{array}{l}
{\langle\psi_R (p)\bar{\psi}_R (p)\rangle = \dfrac{i(\slashed{p}+m_\psi)}{p^2-m_\psi^2}\,+\,\text{Regular terms},
}\\[4pt]
{\langle\phi_R (p)\phi_R (-p)\rangle = \dfrac{i}{p^2-M_\phi^2}\,+\,\text{Regular terms}.}
\end{array}
\end{equation*}

The fields $\psi_R$ and $\phi_R$ are obtained from the bare fields, $\psi$ and $\phi$ --the fields in (\ref{classactionGR})-- by introducing the following wave function renormalizations
\begin{equation}
\begin{array}{l}
{\psi=Z_\psi^{1/2}\,\psi_R,\quad \phi=Z_\phi^{1/2}\,\phi_R,}\\[4pt]
{Z_\psi=1+\delta Z_\psi,\quad \delta Z_\psi=\Sigma_ 2(p^2=m_\psi^2)+ 2 m_\psi^2\Sigma^{\,'}_1(p^2=m_\psi^2),\quad\Sigma^{\,'}_1(p^2)\equiv\dfrac{d \ }{dp^2}\Sigma_1(p^2), }\\[4pt]
{Z_\phi=1+\delta Z_\phi,\quad \delta Z_\phi=i\Gamma^{\,'}_{\phi\phi}(p^2=M_\phi^2),\quad \Gamma^{\,'}_{\phi\phi}(p^2)\equiv \dfrac{d \ }{dp^2}\Gamma_{\phi\phi}(p^2).}
\end{array}
\label{wavefunctionren}
\end{equation}

The reader should bear in mind that in the defining $Z_m$, $Z_\psi$, $Z_M$ and $Z_\phi$ in terms of $i\Gamma_{\psi\bar{\psi}}(p,\kappa)$ and $ i\Gamma_{\phi \phi}(p;\kappa)$, we have taken into account that we are working at the one-loop level.

Considering (\ref{contributionsone}), (\ref{wavefunctionren}) and the definitions in (\ref{zetamass}), one obtains
\begin{equation}
\begin{array}{l}
{\delta Z_\psi=\dfrac{1}{16\pi^2\epsilon}\left[\dfrac{1}{2}g^2+\kappa^2 m^2_\psi \left(\dfrac{5}{8}+\left[\dfrac{1}{2}+\alpha\right]\right)\right]+\text{UV finite contributions},}\\[8pt]
{\delta Z_\phi=\dfrac{1}{16\pi^2\epsilon}\left[2g^2 +\kappa^2 M^2_\phi\left(1+\left[\dfrac{1}{2}+\alpha\right]\right)\right]+\text{UV finite contributions},}
\end{array}
\label{wavefuncresults}
\end{equation}
where $n=4+2\epsilon$, $n$ being the dimension of spacetime in Dimensional Regularization. The bits of $\delta Z_\psi$ and $\delta Z_\phi$ in (\ref{wavefuncresults}) that are independent of $\kappa$ are the usual ones that can be found in textbooks.

The wave function renormalizations in (\ref{wavefunctionren}) and (\ref{wavefuncresults}) give rise to a vertex counterterm diagrammatically represented by the diagram in Figure \ref{figure5}, whose value reads
\begin{equation}
\begin{array}{ll}
-i g (\delta Z_\psi+\frac{1}{2}\delta Z_\phi)&=-\dfrac{ig}{16\pi^2\epsilon}\left[\dfrac{3}{2} g^2 +\kappa^2 m^2_\psi\left(\dfrac{5}{8}+\left[\dfrac{1}{2}+\alpha\right]\right)+\kappa^2 M^2_\phi\left(1+\left[\dfrac{1}{2}+\alpha\right]\right)\right]\\[7pt]
&+\text{UV finite contributions}.
\end{array}
\label{waverencount}
\end{equation}

\begin{figure}[h]
\centering
\begin{fmffile}{yukawacounter}
	\begin{fmfgraph*}(100,110)
		\fmfbottom{i,o}
		\fmftop{t}
		\fmf{scalar,tension=2,label=$p$}{t,v1}
		\fmf{fermion,tension=1,label=$p_1$}{v1,i}
		\fmf{fermion,tension=1,label=$p_2$}{o,v1}
		\fmfv{decor.shape=circle,decor.filled=empty,
			decor.size=(.24w)}{v1}
		\fmfv{l=\scalebox{3}{$\mathbf {\times}$},label.dist=0}{v1}
			\end{fmfgraph*}
\end{fmffile}
\caption{Counterterm $g\kappa^2$.}
\label{figure5}
\end{figure}

\setcounter{table}{5} 

\begin{table}[ht!]
	\centering
	\begin{tabular}{m{6cm} m{2cm} m{6cm}}
		\begin{fmffile}{yuk1}
			\begin{fmfgraph*}(160,90)
				\fmfleft{i1,i2}
				\fmfright{f1,f2}
					\fmf{fermion,tension=1,label=$p_2$,l.side=left}{i1,v2}			
				\fmf{fermion,tension=1,label=$p_1$,l.side=left}{v2,i2}
				\fmf{scalar,tension=0.5,label=$Q$,l.side=right}{v3,v2}
				\fmf{fermion,tension=1,label=$p_2'$}{f1,v3}
				\fmf{fermion,tension=1,label=$p_1'$,l.side=right}{v3,f2}
				\fmfblob{0.8cm}{v2}
			\end{fmfgraph*}
		\end{fmffile}&\centering $+$&\begin{fmffile}{yuk2}
		\begin{fmfgraph*}(160,90)
			\fmfleft{i1,i2}
			\fmfright{f1,f2}
			\fmf{fermion,tension=1,label=$p_2$,l.side=left}{i1,v2}			
			\fmf{fermion,tension=1,label=$p_1$}{v2,i2}
			\fmf{scalar,tension=0.5,label=$Q$,l.side=right}{v3,v2}
			\fmf{fermion,tension=1,label=$p_2'$}{f1,v3}
			\fmf{fermion,tension=1,label=$p_1'$,l.side=right}{v3,f2}
			\fmfv{decor.shape=circle,decor.filled=empty,
				decor.size=(.15w)}{v2}
			\fmfv{l=\scalebox{3}{$\mathbf {\times}$},label.dist=0}{v2}
		\end{fmfgraph*}
	\end{fmffile}\\
\begin{fmffile}{yuk3}
	\begin{fmfgraph*}(160,90)
		\fmfleft{i1,i2}
		\fmfright{f1,f2}
		\fmf{fermion,tension=1,label=$p_2$,l.side=left}{i1,v2}			
		\fmf{fermion,tension=1,label=$p_1$,l.side=left}{v2,i2}
		\fmf{scalar,tension=0.5,label=$Q$,l.side=right}{v3,v2}
		\fmf{fermion,tension=1,label=$p_2'$}{f1,v3}
		\fmf{fermion,tension=1,label=$p_1'$,l.side=right}{v3,f2}
		\fmfblob{0.8cm}{v3}
	\end{fmfgraph*}
\end{fmffile}&\centering $+$&\begin{fmffile}{yuk4}
	\begin{fmfgraph*}(160,90)
		\fmfleft{i1,i2}
		\fmfright{f1,f2}
		\fmf{fermion,tension=1,label=$p_2$,l.side=left}{i1,v2}			
		\fmf{fermion,tension=1,label=$p_1$,l.side=right}{v2,i2}
		\fmf{scalar,tension=0.5,label=$Q$,l.side=right}{v3,v2}
		\fmf{fermion,tension=1,label=$p_2'$}{f1,v3}
		\fmf{fermion,tension=1,label=$p_1'$,l.side=right}{v3,f2}
		\fmfv{decor.shape=circle,decor.filled=empty,
			decor.size=(.15w)}{v3}
		\fmfv{l=\scalebox{3}{$\mathbf {\times}$},label.dist=0}{v3}
	\end{fmfgraph*}
\end{fmffile}
\end{tabular}
	\caption{S matrix contributions.}
\label{figure6}
\end{table}
\setcounter{figure}{6}

\begin{table}[ht!]
	\centering
	\begin{tabular}{m{6cm} m{2cm} m{6cm}}
		\begin{fmffile}{yuk1cross}
			\begin{fmfgraph*}(160,90)
				\fmfleft{i1,i2}
				\fmfright{f1,f2}
				\fmf{fermion,tension=1,label=$p_2$,l.side=left}{i1,v2}			
				\fmf{fermion,tension=1,label=$p_1'$,l.side=left}{v2,i2}
				\fmf{scalar,tension=0.5,label=$Q$,l.side=right}{v3,v2}
				\fmf{fermion,tension=1,label=$p_2'$}{f1,v3}
				\fmf{fermion,tension=1,label=$p_1$,l.side=right}{v3,f2}
				\fmfblob{0.8cm}{v2}
			\end{fmfgraph*}
		\end{fmffile}&\centering $+$&\begin{fmffile}{yuk2cross}
			\begin{fmfgraph*}(160,90)
				\fmfleft{i1,i2}
				\fmfright{f1,f2}
				\fmf{fermion,tension=1,label=$p_2$,l.side=left}{i1,v2}			
				\fmf{fermion,tension=1,label=$p_1'$}{v2,i2}
				\fmf{scalar,tension=0.5,label=$Q$,l.side=right}{v3,v2}
				\fmf{fermion,tension=1,label=$p_2'$}{f1,v3}
				\fmf{fermion,tension=1,label=$p_1$,l.side=right}{v3,f2}
				\fmfv{decor.shape=circle,decor.filled=empty,
					decor.size=(.15w)}{v2}
				\fmfv{l=\scalebox{3}{$\mathbf {\times}$},label.dist=0}{v2}
			\end{fmfgraph*}
		\end{fmffile}\\
		\begin{fmffile}{yuk3cross}
			\begin{fmfgraph*}(160,90)
				\fmfleft{i1,i2}
				\fmfright{f1,f2}
				\fmf{fermion,tension=1,label=$p_2$,l.side=left}{i1,v2}			
				\fmf{fermion,tension=1,label=$p_1'$,l.side=left}{v2,i2}
				\fmf{scalar,tension=0.5,label=$Q$,l.side=right}{v3,v2}
				\fmf{fermion,tension=1,label=$p_2'$}{f1,v3}
				\fmf{fermion,tension=1,label=$p_1$,l.side=right}{v3,f2}
				\fmfblob{0.8cm}{v3}
			\end{fmfgraph*}
		\end{fmffile}&\centering $+$&\begin{fmffile}{yuk4cross}
			\begin{fmfgraph*}(160,90)
				\fmfleft{i1,i2}
				\fmfright{f1,f2}
				\fmf{fermion,tension=1,label=$p_2$,l.side=left}{i1,v2}			
				\fmf{fermion,tension=1,label=$p_1'$,l.side=right}{v2,i2}
				\fmf{scalar,tension=0.5,label=$Q$,l.side=right}{v3,v2}
				\fmf{fermion,tension=1,label=$p_2'$}{f1,v3}
				\fmf{fermion,tension=1,label=$p_1$,l.side=right}{v3,f2}
				\fmfv{decor.shape=circle,decor.filled=empty,
					decor.size=(.15w)}{v3}
				\fmfv{l=\scalebox{3}{$\mathbf {\times}$},label.dist=0}{v3}
			\end{fmfgraph*}
		\end{fmffile}
	\end{tabular}
	\caption{Crossing S matrix contributions.}
\label{figure7}
\end{table}
\setcounter{figure}{7}
Using the value of $i\Gamma^{(GR)}_{\psi\bar{\psi}\phi}(p_1,p_2;\kappa)$, displayed in (\ref{contributionsone}), and
the counterterm in (\ref{waverencount}), one obtains the following expression for the UV divergent General Relativity  contribution  to the S matrix coming from the sum of all the diagrams in Figure \ref{figure6}:
\begin{equation}
\begin{array}{l}
{(\bar{u}(p_1)\cdot u(p_2))(\bar{u}(p^{'}_1)\cdot u(p^{'}_2))\times}\\[4pt]
\times{\left(\dfrac{-i}{16\pi^2\epsilon}\right)g^2\kappa^2\left[\dfrac{1}{2}M^2_\phi+\dfrac{3}{2}m^2_\psi-\dfrac{1}{8}Q^2+\left[\frac{1}{2}+\alpha\right]\left(Q^2-M^2_\phi\right)\right]
\dfrac{i}{Q^2-M^2_\phi},}
\end{array}
\label{propagatingHiggs}
\end{equation}
where $Q=p_1-p_2$. Bear in mind that the blob with slanted lines in diagrams of Figure \ref{figure6} represents the sum of all the diagrams in
Figure \ref{figure3}, i.e. $i\Gamma^{(GR)}_{\psi\bar{\psi}\phi}(p_1,p_2;\kappa)$.

The crossing diagrams in Figure \ref{figure7} yields the following UV divergent General Relativity contribution to the S matrix:
\begin{equation}
\begin{array}{l}
{-(\bar{u}(p_1)\cdot u(p^{'}_2))(\bar{u}(p^{'}_1)\cdot u(p_2))\times}\\[4pt]
{\left(\dfrac{-i}{16\pi^2\epsilon}\right)g^2\kappa^2\left[\dfrac{1}{2}M^2_\phi+\frac{3}{2}m^2_\psi-\dfrac{1}{8}\tilde{Q}^2+\left[\dfrac{1}{2}+\alpha\right](\tilde{Q}^2-M^2_\phi)\right]
\dfrac{i}{\tilde{Q}^2-M^2_\phi},}
\end{array}
\label{CrossingpropHiggs}
\end{equation}
where $\tilde{Q}=p^{'}_1-p_2$.

Let us denote by Box8a, Box8b and Box8c the UV divergent General Relativity UV divergent contributions to the S matrix coming from the diagrams a), b) an c) in Figure \ref{figure8}, respectively. A lengthy computation yields at the following simple expressions
\begin{equation*}
\begin{array}{l}
{\text{Box8a}= \Big(\dfrac{1}{16\pi^2\epsilon}\Big)\kappa^2 g^2 \left(-\dfrac{3}{8}\right)(\bar{u}(p_1)\cdot u(p_2))(\bar{u}(p^{'}_1)\cdot u(p^{'}_2))},\\[4pt]
{\text{Box8b}=\Big(\dfrac{1}{16\pi^2\epsilon}\Big)\kappa^2 g^2 \left(\dfrac{3}{2}\right)(\bar{u}(p_1)\cdot u(p_2))(\bar{u}(p^{'}_1)\cdot u(p^{'}_2))},\\[4pt]
{\text{Box8c}=\Big(\dfrac{1}{16\pi^2\epsilon}\Big)\kappa^2 g^2\Big[-1-[\dfrac{1}{2}+\alpha]\Big](\bar{u}(p_1)\cdot u(p_2))(\bar{u}(p^{'}_1)\cdot u(p^{'}_2)).}
\end{array}
\end{equation*}

\begin{figure}[h!]
	\centering
	\begin{subfigure}{0.3\linewidth}
		\begin{fmffile}{BOX1-4}
			\begin{fmfgraph*}(160,90)
				\fmfleft{i1,i2}
				\fmfright{f1,f2}
				\fmf{fermion,tension=1,label=$p_2$}{i1,v1}
				\fmf{fermion,tension=1,label=$p_2+k$,l.side=left}{v1,v2}
				\fmf{fermion,tension=0.5,label=$p_1$}{v2,i2}
				\fmf{scalar,tension=0.5,label=$k+Q$,l.side=left}{v2,v3}
				\fmf{wiggly, tension=0, label=$k$,l.side=right}{v1,v4}
				\fmf{fermion,tension=1,label=$p_2'$}{f1,v4}
				\fmf{fermion,tension=1,label=$p_2'-k$,l.side=right}{v4,v3}
				\fmf{fermion,tension=0.5,label=$p_1'$,l.side=right}{v3,f2}
			\end{fmfgraph*}
		\end{fmffile}
		\caption{+3 permutations}
	\end{subfigure}\hfill
	\begin{subfigure}{0.3\linewidth}
		\begin{fmffile}{BOX5-8}
			\begin{fmfgraph*}(160,90)
				\fmfleft{i1,i2}
				\fmfright{f1,f2}
				\fmf{fermion,tension=0.5,label=$p_2$}{i1,v2}
				\fmf{fermion,tension=0.5,label=$p_1$}{v2,i2}
				\fmf{scalar,tension=0.5,label=$k+Q$,l.side=left}{v2,v3}
				\fmf{wiggly, right=0.2,tension=0, label=$k$,l.side=right}{v2,v4}
				\fmf{fermion,tension=1,label=$p_2'$}{f1,v4}
				\fmf{fermion,tension=1,label=$p_2'-k$,l.side=right}{v4,v3}
				\fmf{fermion,tension=0.5,label=$p_1'$}{v3,f2}
			\end{fmfgraph*}
		\end{fmffile}
		\caption{+3 permutations}
	\end{subfigure}\hfill
	\begin{subfigure}{0.3\linewidth}
		\begin{fmffile}{BOX9}
			\begin{fmfgraph*}(160,90)
				\fmfleft{i1,i2}
				\fmfright{f1,f2}
				\fmf{fermion,tension=0.5,label=$p_2$}{i1,v2}
				\fmf{fermion,tension=0.5,label=$p_1$,l.side=left}{v2,i2}
				\fmf{scalar,tension=0.5,label=$Q$,l.side=left}{v2,v3}
				\fmf{wiggly, tension=0.1,right, label=$k$,l.side=right}{v2,v3}
				\fmf{fermion,tension=0.5,label=$p_2'$}{f1,v3}
				\fmf{fermion,tension=0.5,label=$p_1'$,l.side=right}{v3,f2}
			\end{fmfgraph*}
		\end{fmffile}
		\caption{}
	\end{subfigure}
	\caption{Box diagrams.}
	\label{figure8}
\end{figure}
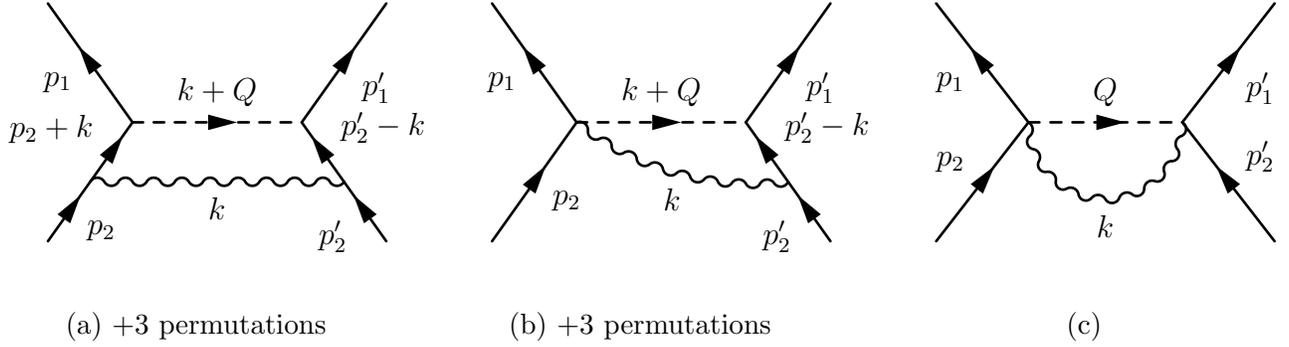

\begin{figure}[h!]
	\centering
	\begin{subfigure}{0.3\linewidth}
		\begin{fmffile}{BOX1-4cross}
			\begin{fmfgraph*}(160,90)
				\fmfleft{i1,i2}
				\fmfright{f1,f2}
				\fmf{fermion,tension=1,label=$p_2$}{i1,v1}
				\fmf{fermion,tension=1,label=$p_2+k$,l.side=left}{v1,v2}
				\fmf{fermion,tension=0.5,label=$p_1'$}{v2,i2}
				\fmf{scalar,tension=0.5,label=$k+Q$,l.side=left}{v2,v3}
				\fmf{wiggly, tension=0, label=$k$,l.side=right}{v1,v4}
				\fmf{fermion,tension=1,label=$p_2'$}{f1,v4}
				\fmf{fermion,tension=1,label=$p_2'-k$,l.side=right}{v4,v3}
				\fmf{fermion,tension=0.5,label=$p_1$}{v3,f2}
			\end{fmfgraph*}
		\end{fmffile}
		\caption{+3 permutations}
	\end{subfigure}\hfill
	\begin{subfigure}{0.3\linewidth}
		\begin{fmffile}{BOX5-8cross}
			\begin{fmfgraph*}(160,90)
				\fmfleft{i1,i2}
				\fmfright{f1,f2}
				\fmf{fermion,tension=0.5,label=$p_2$}{i1,v2}
				\fmf{fermion,tension=0.5,label=$p_1'$}{v2,i2}
				\fmf{scalar,tension=0.5,label=$k+Q$,l.side=left}{v2,v3}
				\fmf{wiggly, right=0.2,tension=0, label=$k$,l.side=right}{v2,v4}
				\fmf{fermion,tension=1,label=$p_2'$}{f1,v4}
				\fmf{fermion,tension=1,label=$p_2'-k$,l.side=right}{v4,v3}
				\fmf{fermion,tension=0.5,label=$p_1$}{v3,f2}
			\end{fmfgraph*}
		\end{fmffile}
		\caption{+3 permutations}
	\end{subfigure}\hfill
	\begin{subfigure}{0.3\linewidth}
		\begin{fmffile}{BOX9cross}
			\begin{fmfgraph*}(160,90)
				\fmfleft{i1,i2}
				\fmfright{f1,f2}
				\fmf{fermion,tension=0.5,label=$p_2$}{i1,v2}
				\fmf{fermion,tension=0.5,label=$p_1'$,l.side=left}{v2,i2}
				\fmf{scalar,tension=0.5,label=$Q$,l.side=left}{v2,v3}
				\fmf{wiggly, tension=0.1,right, label=$k$,l.side=right}{v2,v3}
				\fmf{fermion,tension=0.5,label=$p_2'$}{f1,v3}
				\fmf{fermion,tension=0.5,label=$p_1$,l.side=right}{v3,f2}
			\end{fmfgraph*}
		\end{fmffile}
		\caption*{}
	\end{subfigure}
	\caption{Box diagrams (crossing).}
	\label{figure9}
\end{figure}
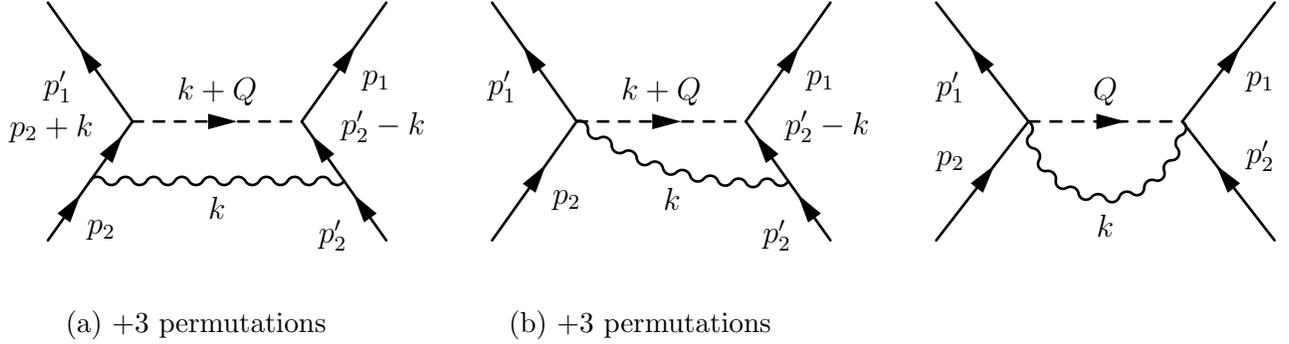

By adding Box8a, Box8b and Box8c above, one obtains
\begin{equation}
\Big(\dfrac{1}{16\pi^2\epsilon}\Big)\kappa^2 g^2\left(\dfrac{1}{8}-\left(\dfrac{1}{2}+\a\right)\right)(\bar{u}(p_1)\cdot u(p_2))(\bar{u}(p^{'}_1)\cdot u(p^{'}_2)).
\label{Boxdia}
\end{equation}

Obviously, the sum of the crossing diagrams in Figure \ref{figure9} yields the following UV contribution to the S matrix for General Relativity
\begin{equation}
-\Big(\dfrac{1}{16\pi^2\epsilon}\Big)\kappa^2 g^2\left(\dfrac{1}{8}-\left(\dfrac{1}{2}+\a\right)\right)(\bar{u}(p^{'}_1)\cdot u(p_2))(\bar{u}(p_1)\cdot u(p^{'}_2)).
\label{crossboxdiag}
\end{equation}

Adding the expressions in (\ref{propagatingHiggs}), (\ref{CrossingpropHiggs}), (\ref{Boxdia}) and (\ref{crossboxdiag}), one obtains the following result
\begin{equation}
\begin{array}{l}
{(\bar{u}(p_1)\cdot u(p_2))(\bar{u}(p^{'}_1)\cdot u(p^{'}_2))
\left(\dfrac{1}{16\pi^2\epsilon}\right)g^2\kappa^2\left\{\left[\dfrac{3}{8}M^2_\phi+\dfrac{3}{2}m^2_\psi\right]
\dfrac{1}{Q^2-M^2_\phi}\right\}}-\\[4pt]
{-(\bar{u}(p_1)\cdot u(p^{'}_2))(\bar{u}(p^{'}_1)\cdot u(p_2))
\left(\dfrac{1}{16\pi^2\epsilon}\right)g^2\kappa^2\left\{\left[\dfrac{3}{8}M^2_\phi+\dfrac{3}{2}m^2_\psi\right]
\dfrac{1}{\tilde{Q}^2-M^2_\phi}\right\},}
\end{array}
\label{resultfig6789}
\end{equation}
where $Q=p_1-p_2$ and $\tilde{Q}=p^{'}_1-p_2$. Let us stress that the dependence on the gauge parameter $\alpha$  goes away upon summing over all diagrams in Figures \ref{figure6} to \ref{figure9}.

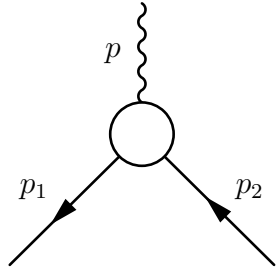
\begin{figure}[h!]
	\centering
	\begin{fmffile}{yukawacounter2}
		\begin{fmfgraph*}(100,110)
			\fmfbottom{i,o}
			\fmftop{t}
			\fmf{wiggly,tension=2,label=$p$}{t,v1}
			\fmf{fermion,tension=1,label=$p_1$}{v1,i}
			\fmf{fermion,tension=1,label=$p_2$}{o,v1}
			\fmfv{decor.shape=circle,decor.filled=empty,
				decor.size=(.24w)}{v1}
		\end{fmfgraph*}
	\end{fmffile}
	\caption{Counterterm $ g^2\kappa$.}
	\label{figure10}
\end{figure}

Let us introduce now the counterterm vertex in Figure \ref{figure10}, which comes from renormalization produced by the constants $Z_\psi$ and $Z_m$ in Minkowski spacetime applied to the energy-momentum tensor in (\ref{EMtensor}):
\begin{equation}
\begin{array}{l}
{i\Gamma^{(ct)}_{\psi\bar{\psi} h_{\mu\nu}}(p_1,p_2;\kappa)=}\\[4pt]
{-i\dfrac{\kappa}{2}\Big\{\delta Z^{(Mink)}_\psi\Big[\dfrac{1}{4}\big(\gamma^\mu (p_1+p_2)^\nu+\gamma^\nu (p_1+p_2)^\mu\big)-\dfrac{1}{2}\eta^{\mu\nu}
\big((\slashed{p}_1+
\slashed{p}_2)- 2m \big)\Big]+\delta Z^{(Mink)}_m \eta^{\mu\nu} m\Big\},}\end{array}
\label{newcounterterm}
\end{equation}
where 
\begin{equation*}{\phantom{-i\dfrac{\kappa}{2}\Big\{}\delta Z^{(Mink)}_\psi = \dfrac{1}{2} g^2 \Big(\dfrac{1}{16\pi\epsilon}\Big),\quad\delta Z^{(Mink)}_m = -\dfrac{3}{2} g^2 \Big(\dfrac{1}{16\pi\epsilon}\Big).}\end{equation*}

Taking into account the expressions presented in (\ref{contributionsone}) and (\ref{newcounterterm}), one concludes that

\begin{equation}
i\Gamma^{(GR)}_{\psi\bar{\psi} h_{\mu\nu}}(p_1,p_2;\kappa)+ i\Gamma^{(ct)}_{\psi\bar{\psi} h_{\mu\nu}}(p_1,p_2;\kappa)\,=\, 0.
\label{magical}
\end{equation}

Hence, at one-loop and $g^2\kappa$ order, the vertex $\bar{\psi}\psi h_{\mu\nu}$ is UV finite and, as a result, there is no UV divergent contribution to the S matrix coming from the sum of the diagrams in Figure \ref{figure11}. Likewise for the sum of the diagrams in Figure 12. Note that the black blob in Figures \ref{figure11} and \ref{figure12} represents $i\Gamma^{(GR)}_{\psi\bar{\psi} h_{\mu\nu}}(p_1,p_2;\kappa)$ in (\ref{contributionsone}).

The cancellation of UV divergences in (\ref{magical}) is an addition to the list of surprising UV cancellations that occur when gravity is coupled to matter: See \cite{Bace:1980uk, Ebert:2007gf} and references therein.

Summarizing, the UV divergent contribution the S matrix element of the $fermion + fermion \rightarrow fermion + fermion$ scattering is given by (\ref{resultfig6789}) at one-loop and $g^2\kappa^2$ order,
due to the cancellation of UV divergences in (\ref{magical}).

\setcounter{table}{10} 

\begin{table}[h!]
	\centering
	\begin{tabular}{m{6cm} m{2cm} m{6cm}}
		\begin{fmffile}{yukg1}
			\begin{fmfgraph*}(160,90)
				\fmfleft{i1,i2}
				\fmfright{f1,f2}
				\fmf{fermion,tension=1,label=$p_2$,l.side=left}{i1,v2}			
				\fmf{fermion,tension=1,label=$p_1$,l.side=left}{v2,i2}
				\fmf{wiggly,tension=0.5,label=$Q$,l.side=right}{v3,v2}
				\fmf{fermion,tension=1,label=$p_2'$}{f1,v3}
				\fmf{fermion,tension=1,label=$p_1'$,l.side=right}{v3,f2}
				\fmfv{decor.shape=circle,decor.filled=full,	decor.size=(.15w)}{v2}
			\end{fmfgraph*}
		\end{fmffile}&\centering $+$&\begin{fmffile}{yukg2}
			\begin{fmfgraph*}(160,90)
				\fmfleft{i1,i2}
				\fmfright{f1,f2}
				\fmf{fermion,tension=1,label=$p_2$,l.side=left}{i1,v2}			
				\fmf{fermion,tension=1,label=$p_1$}{v2,i2}
				\fmf{wiggly,tension=0.5,label=$Q$,l.side=right}{v3,v2}
				\fmf{fermion,tension=1,label=$p_2'$}{f1,v3}
				\fmf{fermion,tension=1,label=$p_1'$,l.side=right}{v3,f2}
				\fmfv{decor.shape=circle,decor.filled=empty,
					decor.size=(.15w)}{v2}
			\end{fmfgraph*}
		\end{fmffile}\\
		\begin{fmffile}{yukg3}
			\begin{fmfgraph*}(160,90)
				\fmfleft{i1,i2}
				\fmfright{f1,f2}
				\fmf{fermion,tension=1,label=$p_2$,l.side=left}{i1,v2}			
				\fmf{fermion,tension=1,label=$p_1$,l.side=left}{v2,i2}
				\fmf{wiggly,tension=0.5,label=$Q$,l.side=right}{v3,v2}
				\fmf{fermion,tension=1,label=$p_2'$}{f1,v3}
				\fmf{fermion,tension=1,label=$p_1'$,l.side=right}{v3,f2}
				\fmfv{decor.shape=circle,decor.filled=full,	decor.size=(.15w)}{v3}
			\end{fmfgraph*}
		\end{fmffile}&\centering $+$&\begin{fmffile}{yukg4}
			\begin{fmfgraph*}(160,90)
				\fmfleft{i1,i2}
				\fmfright{f1,f2}
				\fmf{fermion,tension=1,label=$p_2$,l.side=left}{i1,v2}			
				\fmf{fermion,tension=1,label=$p_1$,l.side=right}{v2,i2}
				\fmf{wiggly,tension=0.5,label=$Q$,l.side=right}{v3,v2}
				\fmf{fermion,tension=1,label=$p_2'$}{f1,v3}
				\fmf{fermion,tension=1,label=$p_1'$,l.side=right}{v3,f2}
				\fmfv{decor.shape=circle,decor.filled=empty,
					decor.size=(.15w)}{v3}
			\end{fmfgraph*}
		\end{fmffile}
	\end{tabular}
	\caption{S matrix contributions.}
	\label{figure11}
\end{table}
\setcounter{figure}{11}

\begin{table}[h!]
	\centering
	\begin{tabular}{m{6cm} m{2cm} m{6cm}}
		\begin{fmffile}{yuk1gcross}
			\begin{fmfgraph*}(160,90)
				\fmfleft{i1,i2}
				\fmfright{f1,f2}
				\fmf{fermion,tension=1,label=$p_2$,l.side=left}{i1,v2}			
				\fmf{fermion,tension=1,label=$p_1'$,l.side=left}{v2,i2}
				\fmf{wiggly,tension=0.5,label=$Q$,l.side=right}{v3,v2}
				\fmf{fermion,tension=1,label=$p_2'$}{f1,v3}
				\fmf{fermion,tension=1,label=$p_1$,l.side=right}{v3,f2}
				\fmfv{decor.shape=circle,decor.filled=full,	decor.size=(.15w)}{v2}
			\end{fmfgraph*}
		\end{fmffile}&\centering $+$&\begin{fmffile}{yuk2gcross}
			\begin{fmfgraph*}(160,90)
				\fmfleft{i1,i2}
				\fmfright{f1,f2}
				\fmf{fermion,tension=1,label=$p_2$,l.side=left}{i1,v2}			
				\fmf{fermion,tension=1,label=$p_1'$}{v2,i2}
				\fmf{wiggly,tension=0.5,label=$Q$,l.side=right}{v3,v2}
				\fmf{fermion,tension=1,label=$p_2'$}{f1,v3}
				\fmf{fermion,tension=1,label=$p_1$,l.side=right}{v3,f2}
				\fmfv{decor.shape=circle,decor.filled=empty,
					decor.size=(.15w)}{v3}
			\end{fmfgraph*}
		\end{fmffile}\\
		\begin{fmffile}{yuk3gcross}
			\begin{fmfgraph*}(160,90)
				\fmfleft{i1,i2}
				\fmfright{f1,f2}
				\fmf{fermion,tension=1,label=$p_2$,l.side=left}{i1,v2}			
				\fmf{fermion,tension=1,label=$p_1'$,l.side=left}{v2,i2}
				\fmf{wiggly,tension=0.5,label=$Q$,l.side=right}{v3,v2}
				\fmf{fermion,tension=1,label=$p_2'$}{f1,v3}
				\fmf{fermion,tension=1,label=$p_1$,l.side=right}{v3,f2}
				\fmfv{decor.shape=circle,decor.filled=full,	decor.size=(.15w)}{v3}
			\end{fmfgraph*}
		\end{fmffile}&\centering $+$&\begin{fmffile}{yuk4gcross}
			\begin{fmfgraph*}(160,90)
				\fmfleft{i1,i2}
				\fmfright{f1,f2}
				\fmf{fermion,tension=1,label=$p_2$,l.side=left}{i1,v2}			
				\fmf{fermion,tension=1,label=$p_1'$,l.side=right}{v2,i2}
				\fmf{wiggly,tension=0.5,label=$Q$,l.side=right}{v3,v2}
				\fmf{fermion,tension=1,label=$p_2'$}{f1,v3}
				\fmf{fermion,tension=1,label=$p_1$,l.side=right}{v3,f2}
				\fmfv{decor.shape=circle,decor.filled=empty,
					decor.size=(.15w)}{v3}
			\end{fmfgraph*}
		\end{fmffile}
	\end{tabular}
	\caption{Crossing S matrix contributions.}
	\label{figure12}
\end{table}

\subsection{The Unimodular Gravity case}

When Unimodular Gravity is the theory of quantum gravity the diagrams to be computed are the same as those in the General Relativity case --ie, diagrams in Figures \ref{figure1} to \ref{figure12}, with the proviso that the internal graviton line represents the propagator --given in (\ref{tracelessprop})-- of the field $\hat{h}_{\mu\nu}$ --this is the field which couples to the Energy-momentum tensor, see (\ref{Sphi4expan}).

Let us denote by $i\Gamma^{(UG)}_{\phi\phi}(p^2,\kappa)$ and  $i\Gamma^{(UG)}_{\psi\bar{\psi}}(p;\kappa)$ the UV divergent part of the diagrams in Figures \ref{figure1} and \ref{figure2}. Let $i\Gamma^{(UG)}_{\psi\bar{\psi}\phi}(p_1,p_2;\kappa)$ be UV divergent contribution coming from the sum of all diagrams in Figure \ref{figure3}, and, finally, let $i\Gamma^{(UG)}_{\psi\bar{\psi}\hat{h}_{\mu\nu}}(p_1,p_2;\kappa)$ stand for the UV divergent contribution obtained by summing all diagrams in Figure \ref{figure4}. We have

\begin{equation}
\begin{array}{l}
{i\Gamma^{(UG)}_{\phi\phi}(p^2,\kappa)= 0,}\\[4pt]
{i\Gamma^{(UG)}_{\psi\bar{\psi}}(p;\kappa)=
\Big(\dfrac{-i}{16\pi^2\epsilon}\Big)\kappa^2 \left(\dfrac{3}{8}p^2 m-\dfrac{5}{16}p^2\slashed{p}+\dfrac{3}{16}m^2\slashed{p}\right),
}\\[4pt]
{i\Gamma^{(UG)}_{\psi\bar{\psi}\phi}(p_1,p_2;\kappa)=g\kappa^2 \left(-\dfrac{i}{16\pi^2\epsilon}\right)\Big( \dfrac{9}{16}(p_1^2+p_2^2)-\dfrac{3}{8} p_1\cdot p_2+\dfrac{3}{16}m(\slashed{p}_1+\slashed{p}_2)-\dfrac{3}{8}\slashed{p}_1\slashed{p}_2\Big),}\\[4pt]
{i\Gamma^{(UG)}_{\psi\bar{\psi}\hat{ h}_{\mu\nu}}(p_1,p_2;\kappa)=\kappa g^2\Big(\dfrac{-i}{16\pi^2\epsilon}\Big)\Big\{\Big[\dfrac{1}{8}(\slashed{p}_1+\slashed{p}_2)+\frac{1}{2}m\Big]\eta^{\mu\nu}-\dfrac{1}{16}(p_1+p_2)^{\mu}\gamma^\nu-\dfrac{1}{16}(p_1+p_2)^{\nu}\gamma^\mu\Big\}}
\end{array}
\label{contributionsoneUG}
 \end{equation}

Taking into account the definitions in (\ref{wavefunctionren}) and the results in (\ref{contributionsoneUG}), one concludes, after a little algebra,
that
\begin{equation*}
\begin{array}{l}
{\delta Z_\psi=\dfrac{1}{16\pi^2\epsilon}\frac{1}{2}g^2+\text{UV finite contributions},}\\[4pt]
{\delta Z_\phi=\dfrac{1}{16\pi^2\epsilon}2g^2+\text{UV finite contributions},}
\end{array}
\end{equation*}
i.e. in the Unimodular Gravity case and for the gauge-fixing leading to the propagator in (\ref{tracelessprop}), there are no $g^2\kappa^2$ UV divergent contributions to the wave function renormalizations of the fermion and scalar fields. Hence, unlike in the General Relativity case, the countertem vertex in Figure \ref{figure5} can be set to zero when computing the UV divergent contributions to the S matrix. This very same reasoning applies to those diagrams in Figures \ref{figure6} and \ref{figure7} which involve the vertex in Figure \ref{figure5}.

Let us point out that the blob with slanted lines in Figure \ref{figure6} represents now the function $i\Gamma^{(UG)}_{\psi\bar{\psi}\phi}(p_1,p_2;\kappa)$.  Next, by using the value of $i\Gamma^{(UG)}_{\psi\bar{\psi}\phi}(p_1,p_2;\kappa)$ in (\ref{contributionsoneUG}),  one obtains that the Unimodular Gravity UV divergent contribution to the S matrix coming form the diagrams in Figure \ref{figure6} reads:
\begin{equation}
\begin{array}{l}
{(\bar{u}(p_1)\cdot u(p_2))(\bar{u}(p^{'}_1)\cdot u(p^{'}_2))\times}\\[4pt]
\times{\Big(\dfrac{1}{16\pi^2\epsilon}\Big)g^2\kappa^2\Big[\frac{3}{2}m^2_\psi+\frac{3}{8}Q^2\Big]
\dfrac{1}{Q^2-M^2_\phi},}
\end{array}
\label{propagatingHiggsUG}
\end{equation}
where $Q=p_1-p_2$.

Now, the corresponding contribution coming from the crossing diagrams in Figure 7 runs thus
\begin{equation}
\begin{array}{l}
-{(\bar{u}(p^{'}_1)\cdot u(p_2))(\bar{u}(p_1)\cdot u(p^{'}_2))\times}\\[4pt]
\times{\Big(\dfrac{1}{16\pi^2\epsilon}\Big)g^2\kappa^2\Big[\frac{3}{2}m^2_\psi+\frac{3}{8}\tilde{Q}^2\Big]
\dfrac{1}{\tilde{Q}^2-M^2_\phi},}
\end{array}
\label{crosspropagatingHiggsUG}
\end{equation}
where $\tilde{Q}=p^{'}_1-p_2$.

Let us call Box8aUG, Box8bUG and Box8cUG the sum of the diagrams in Figures 8a, 8b and 8c, respectively, where the graviton line stands for the propagator in (\ref{tracelessprop}). A long computation yields the following simple results for the UV divergent contribution to the S matrix coming from those diagrams:
\begin{equation*}
\begin{array}{l}
{\text{Box8aUG}= \Big(\dfrac{1}{16\pi^2\epsilon}\Big)\kappa^2 g^2 \left(\dfrac{17}{8}\right)(\bar{u}(p_1)\cdot u(p_2))(\bar{u}(p^{'}_1)\cdot u(p^{'}_2))}\\[4pt]
{\text{Box8bUG}=\Big(\dfrac{1}{16\pi^2\epsilon}\Big)\kappa^2 g^2 \left(-\dfrac{3}{2}\right)(\bar{u}(p_1)\cdot u(p_2))(\bar{u}(p^{'}_1)\cdot u(p^{'}_2))}\\[4pt]
{\text{Box8cUG}=\Big(\dfrac{1}{16\pi^2\epsilon}\Big)\kappa^2 g^2(-1)(\bar{u}(p_1)\cdot u(p_2))(\bar{u}(p^{'}_1)\cdot u(p^{'}_2)),}
\end{array}
\end{equation*}
Hence,
\begin{equation}
\text{Box8aUG}+\text{Box8bUG}+\text{Box8cUG}=\Big(\dfrac{1}{16\pi^2\epsilon}\Big)\kappa^2 g^2 \left(-\dfrac{3}{8}\right)(\bar{u}(p_1)\cdot u(p_2))(\bar{u}(p^{'}_1)\cdot u(p^{'}_2))
\label{boxesUG}
\end{equation}

The contribution coming from the crossing diagrams in Figure \ref{figure9} reads
\begin{equation}
\Big(\dfrac{1}{16\pi^2\epsilon}\Big)\kappa^2 g^2 \left(-\dfrac{3}{8}\right)(\bar{u}(p^{'}_1)\cdot u(p_2))(\bar{u}(p_1)\cdot u(p^{'}_2)).
\label{crossboxesUG}
\end{equation}

Let us add all the UV divergent contributions to the S matrix coming from the diagrams in Figures \ref{figure6}, \ref{figure7}, \ref{figure8} and \ref{figure9}. These contributions are given
in (\ref{propagatingHiggsUG}), (\ref{crosspropagatingHiggsUG}), (\ref{boxesUG}) and (\ref{crossboxesUG}); their sum being
\begin{equation}
\begin{array}{l}
{(\bar{u}(p_1)\cdot u(p_2))(\bar{u}(p^{'}_1)\cdot u(p^{'}_2))
\Big(\dfrac{1}{16\pi^2\epsilon}\Big)g^2\kappa^2\Big\{\Big[\dfrac{3}{2}m^2_\psi+\dfrac{3}{8} Q^2-\dfrac{3}{8}(Q^2-M^2_\phi)\Big]
\dfrac{1}{Q^2-M^2_\phi}\Big\}}-\\[4pt]
{-(\bar{u}(p_1)\cdot u(p^{'}_2))(\bar{u}(p^{'}_1)\cdot u(p_2))
\Big(\dfrac{1}{16\pi^2\epsilon}\Big)g^2\kappa^2\Big\{\Big[\dfrac{3}{2}m^2_\psi+\dfrac{3}{8} Q^2-\dfrac{3}{8}(Q^2-M^2_\phi)\Big]
\dfrac{1}{\tilde{Q}^2-M^2_\phi}\Big\}=}\\[4pt]
{(\bar{u}(p_1)\cdot u(p_2))(\bar{u}(p^{'}_1)\cdot u(p^{'}_2))
\Big(\dfrac{1}{16\pi^2\epsilon}\Big)g^2\kappa^2\Big\{\Big[\dfrac{3}{8}M^2_\phi+\frac{3}{2}m^2_\psi\Big]
\dfrac{1}{Q^2-M^2_\phi}\Big\}}-\\[4pt]
{-(\bar{u}(p_1)\cdot u(p^{'}_2))(\bar{u}(p^{'}_1)\cdot u(p_2))
\Big(\dfrac{1}{16\pi^2\epsilon}\Big)g^2\kappa^2\Big\{\Big[\dfrac{3}{8}M^2_\phi+\frac{3}{2}m^2_\psi\Big]
\dfrac{1}{\tilde{Q}^2-M^2_\phi}\Big\},}
\end{array}
\label{resultfig6789UG}
\end{equation}
where $Q=p_1-p_2$ and $\tilde{Q}=p^{'}_1-p_2$

It is plain that (\ref{resultfig6789}) and (\ref{resultfig6789UG}) are equal, i.e., the sum of the UV divergent contributions to the S matrix coming from the diagrams in Figures \ref{figure6}, \ref{figure7}, \ref{figure8} and \ref{figure9} in General Relativity and Unimodular Gravity is the same. Notice, however, that the contribution coming from each diagram is not the same in General Relativity as in Unimodular Gravity.

It is clear --the gravitational field here is a mere spectator-- that the counterterm vertex represented by the diagram in Figure \ref{figure10} has the same value for Unimodular Gravity as for General Relativity, i.e. is given by the expressions in (\ref{newcounterterm}) upon replacing $h_{\mu\nu}$ with $\hat{h}_{\mu\nu}$. Next, notice that $i\Gamma^{(UG)}_{\psi\bar{\psi}\hat{ h}_{\mu\nu}}(p_1,p_2;\kappa)$ in  (\ref{contributionsoneUG}) is equal to $i\Gamma^{(GR)}_{\psi\bar{\psi}\hat{ h}_{\mu\nu}}(p_1,p_2;\kappa)$ in (\ref{contributionsone}). Hence, as in the General Relativity case -see (\ref{magical}), the following cancellation of UV divergences hold
\begin{equation}
i\Gamma^{(UG)}_{\psi\bar{\psi}\hat{h}_{\mu\nu}}(p_1,p_2;\kappa)+ i\Gamma^{(ct)}_{\psi\bar{\psi}\hat{h}_{\mu\nu}}(p_1,p_2;\kappa)\,=\, 0.
\label{magicalUG}
\end{equation}
Thus, the one loop and $g^2\kappa$ order correction to the $\bar{\psi}\psi \hat{h}_{\mu\nu}$ is UV finite.

Using the previous result, one concludes that the sum of the diagrams in Figure \ref{figure11} contains no UV divergent pieces in the Unimodular Gravity case either. Same result for the sum of the diagrams in Figure \ref{figure12}.

In summary, due to the UV cancellation we have just discussed,  only the sum of the diagrams in Figures \ref{figure6}, \ref{figure7}, \ref{figure8} and \ref{figure9} gives, in the Unimodular Gravity case, a UV divergent contribution to the S matrix element of the $fermion + fermion\rightarrow fermion+fermion$ scattering at one-loop and $g^2\kappa^2$ order. Full agreement between Unimodular Gravity and General Relativity has been thus reached.

\section{Summary and discussion}

In this paper we have shown that, at one-loop and $g^2\kappa^2$ order, the UV divergent contribution to the S matrix element of the $fermion + fermion\rightarrow fermion+fermion$ scattering process in the  Yukawa theory coupled to Unimodular Gravity, is same as the corresponding S matrix element when Unimodular  Gravity is replaced with General Relativity --see (\ref{resultfig6789}) and (\ref{resultfig6789UG}) and recall that the sum of the diagrams in Figures \ref{figure11} and \ref{figure12} carries no UV divergence. We should point out that the agreement that we have just mentioned does not hold for each individual Feynman diagram --as can be seen by comparing (\ref{propagatingHiggs}) with (\ref{propagatingHiggsUG}), for instance-- but it unfolds upon adding the contributions coming from classes of  the Feynman diagrams -see (\ref{resultfig6789}) and (\ref{resultfig6789UG}), which yields a  result independent of the gauge parameter. Of course, the gauge symmetries of Unimodular Gravity are not the same --the gauge symmetry of Unimodular Gravity being a constrained one-- as those of General Relativity, and, thus, agreement  between non gauge invariant objects computed in both theories is not to be expected and it does not occur in general --see (\ref{propagatingHiggs}) with (\ref{propagatingHiggsUG}). Hence, that agreement is reached when gauge invariant objects are computed is a far from  trivial result. This non triviality  is  further strengthen by the fact that in Unimodular Gravity the graviton field does not couple neither to the mass terms nor the Yukawa operator in the classical action, whereas the graviton field does couple to those operators in the General Relativity case. This coupling between the graviton field and the mass terms and the Yukawa vertex explains partially the discrepancy between $i\Gamma^{(GR)}_{\psi\bar{\psi}\phi}(p_1,p_2;\kappa)$ and  $i\Gamma^{(UG)}_{\psi\bar{\psi}\phi}(p_1,p_2;\kappa)$ in (\ref{contributionsone}) and (\ref{contributionsoneUG}). Indeed, let us give just one example:  diagram d)  in Figure \ref{figure3} gives a non-vanishing contribution in General Relativity, which reads
\begin{equation*}
\frac{-i}{16\pi^2\epsilon}\kappa^2 g\Big[-\dfrac{5}{8}m^2+\dfrac{3}{16} p_1^2+\dfrac{1}{16}m\slashed{p}_1+[\dfrac{1}{2}+\alpha](-m^2+\dfrac{3}{4}p_1^2+\dfrac{1}{4}m\slashed{p}_1)\Big],
\end{equation*}
whereas it yields a vanishing contribution in the Unimodular Gravity case, for it involves the coupling of the graviton field to the Yukawa vertex.

We have also witnessed --see (\ref{magical}) and (\ref{magicalUG})-- a surprising cancellation of all the UV divergences contributing  to the fermion-fermion-graviton vertex at one loop and $g^2\kappa$ order.
 This cancellation to be added to the list of unexpected UV cancellations when gravity is at work at the quantum level.

A final remark, the results presented in this paper explicitly shows that the beta function of the Yukawa coupling as computed in \cite{Rodigast:2009zj} cannot be used to draw any physically meaningful conclusion, a fact already discussed in \cite{Gonzalez-Martin:2017bvw}. See also \cite{Gonzalez-Martin:2017fwz}, for a similar analysis in the $\lambda \phi^4$ theory.

To carry out some of the computations presented in this paper, we have used Mathematica's xAct \cite{Martingarcia} and Form \cite{Kuipers:2012rf}, independently.

\section{Acknowledgements}
We are grateful to E. Alvarez for illuminating discussions.
This work has received funding from the European Unions Horizon 2020 research and innovation programme under the Marie Sklodowska-Curie grants agreement No 674896 and No 690575. We also have been partially supported by the Spanish MINECO through grants
FPA2014-54154-P and FPA2016-78645-P, COST actions MP1405 (Quantum Structure of Spacetime) and  COST MP1210 (The string theory Universe).  S.G-M acknowledge the support of the Spanish Research Agency (Agencia Estatal de Investigaci\'on) through the grant IFT Centro de Excelencia Severo Ochoa SEV-2016-0597.


\end{document}